# Chemical Vapor Deposition of Ni-doped Iron Germanium Telluride Nanosheets

Matthew Metcalf[a], Jesse Martinez[a], Armella Mushfique[a], Alexander Riou[a], Lutfun Nahar[b], Bamidele Onipede[a], Hui Cai[a,c,*]

[a] Department of Physics, University of California, Merced, Merced CA 95343, United States of America

[b] Department of Physics, Arizona State University, Tempe AZ 85281, United States of America

[c] Materials Sciences Division, Lawrence Berkeley National Laboratory, Berkeley CA 94720, United States of America

* Corresponding author: Hui Cai (hcai6@ucmerced.edu)

## Abstract

Iron germanium telluride (FGT; $Fe_mGe_nTe_2$) compounds have attracted significant interest due to their layered van der Waals structure, relatively high Curie temperature, and tunable magnetic properties. Chemical vapor deposition (CVD) is a particularly promising synthesis route owing to its simplicity, low cost, potential for scalability, and widespread adoption in the semiconductor industry, yet it has not been used previously to synthesize FGT with dopants. Here, we report CVD synthesis of both undoped and Ni-doped FGT nanosheets on $SiO_2$/Si substrates. By adjusting precursor molar ratios, we synthesized Ni-doped FGT with multiple Fe concentrations and a 4% Ni-to-Fe ratio. X-ray photoelectron spectroscopy depth profiling further demonstrates that Ni is present in the bulk of the crystals. This straightforward, low-cost, and CMOS-compatible approach demonstrates a route to Ni-doped FGT nanosheets, establishing a foundation for future characterization of Ni-doped FGT and its potential integration into spintronic devices.

## 1. Introduction

Magnetic van der Waals (vdW) materials are an exciting class of material that can sustain ferromagnetic order down to nanoscale thicknesses, including even the two-dimensional (2D) limit, which was previously believed to be impossible due to the Mermin-Wagner theorem [1,2]. As such, these materials have attracted a great deal of attention over the past decade, offering potential applications in spintronic and magnonic devices [1], on-chip optical communications [6], and quantum computing [6]. Among magnetic vdW materials, Fe-Ge-Te (FGT) ternary compounds have drawn attention due to their itinerant ferromagnetism [7], metallic conductivity [8], perpendicular magnetic anisotropy [9,10], tunable ferromagnetic moment [9,10], and relatively high Curie temperature ($T_C$) [9,10]. These compounds typically have the composition $Fe_mGe_nTe_2$ ($3 \leq m \leq 5$, $1 \leq n \leq 2$), including $Fe_3GeTe_2$ [6,9,11,12], $Fe_{3-x}GeTe_2$ [8], $Fe_4GeTe_2$ [13,14], $Fe_5GeTe_2$ [7], $Fe_{5\pm x}GeTe_2$ [9,10,15,16,2], $Fe_{5-x}Ge_2Te_2$ [10,17,18], as well as related compounds outside this stoichiometry [19,20,21]. As-grown FGT crystals tend to be Fe-deficient [8,17,18,22], so elemental compositions with near-integer values are typically not achieved.

While FGT compounds exhibit relatively high $T_C$ compared to other vdW magnets, it is generally below or slightly above room temperature. For example, $T_C$ for monolayer $Fe_3GeTe_2$ is ~130 K and ~230 K for bulk [7], while $T_C$ for bulk $Fe_5GeTe_2$ is ~260-310 K [23].

Doping with Ni has been demonstrated to increase $T_C$ well above room temperature, reaching as high as ~498 K in Ni-doped $Fe_5GeTe_2$ [24]—high enough for potential use in next-generation nanotechnologies. However, future applications require simple, scalable, and cost-effective synthesis methods that are preferably compatible with CMOS technology [9]. Many methods have been used to synthesize pure and doped FGT, including chemical vapor transport [6,8,13,15,2,18,23,25], solid-state reactions [17,26], molecular beam epitaxy [7,11,16,19], flux-assisted growth [12], pulsed laser deposition [24], and chemical vapor deposition [9]. Several of these methods, while effective for laboratory research, are not well suited for thin-film manufacturing. Solid-state reactions, chemical vapor transport, and flux-assisted growth yield bulk FGT crystals that must be mechanically exfoliated to obtain nanosheets, and therefore do not provide a direct, controllable route to nanosheet synthesis, as mechanical exfoliation is inherently stochastic and low-throughput. Molecular beam epitaxy can grow wafer-scale epitaxial 2D FGT nanosheets, but its slow process rate, high cost, and specialized equipment limit its practicality. Pulsed laser deposition, although less prevalent in industrial settings than methods like chemical vapor deposition, can produce thin, highly textured FGT films with nanoscale thickness and good *c*-axis alignment, rendering it suitable for applications that do not require coherent in-plane crystallinity, atomically sharp interfaces, or high-mobility transport. In contrast, chemical vapor deposition (CVD) offers a promising pathway toward scalable implementation due to its lower cost and widespread use in the semiconductor industry for controlled thin film growth, including on CMOS-compatible $SiO_2$/Si substrates [27].

To date, CVD has been used to produce pure FGT, but not doped FGT. In this short communication, we employ CVD to synthesize undoped and Ni-doped FGT (Ni-FGT) nanosheets on $SiO_2$/Si substrates with varying Fe concentrations. X-ray photoelectron spectroscopy depth profiles show Fe, Ge, Te, and Ni signals throughout the thickness of the crystals, providing evidence that Ni is present within the bulk of the FGT crystal rather than confined to the surface. We note that the structural location of the Ni atoms within the lattice, and any associated changes in magnetic properties, remain subjects for future characterization; the primary aim of this short communication is to demonstrate that CVD provides a straightforward and low-cost route to Ni-FGT, supporting future research and opening a possible path toward its integration into manufacturable spintronic applications.

## 2. Method

### 2.1. Synthesis

The CVD process was based on the cascaded space-confined method reported by Nair et al. [9], with our experimental setup shown in Figure 1(g). Growths were performed in one-inch-diameter quartz tubes mounted in a Thermo Fisher Scientific TF55035A-1 single-zone tube furnace at 580 °C with a 50 °C/min ramp rate, 1 min dwell time, and 1 atm pressure. Prior to heating, the system was purged with Ar (1000 sccm) for 10 min. During growth, a mixture of Ar (95 sccm) and $H_2$ (5 sccm) flowed through the tube. Three precursor mixtures were prepared from $FeCl_2$ (Sigma-Aldrich 372870, 98%), Ge (Sigma-Aldrich 327395, ≥99.999%), KI (Sigma-Aldrich 204102, ≥99.99%), and $NiCl_2$ (Sigma-Aldrich 339350, 98%) with $FeCl_2$:$NiCl_2$:Ge:KI molar ratios of 20:0:10:1, 20:10:10:1, and 16:4:10:1 to synthesize Samples 1, 2, and 3, respectively. A small amount of the precursor mixture was sandwiched between two $SiO_2$/Si substrates and positioned at the tube center, while 50 mg of Te (Sigma-Aldrich 266418, 99.8%) was loaded into a quartz boat positioned 5 cm upstream.

### 2.2. Characterization

Atomic force microscopy (AFM; Asylum MFP-3D Origin with ASYMFMHM-R2 probe), scanning electron microscopy (SEM; Zeiss GeminiSEM 500), and energy-dispersive X-ray spectroscopy (EDX; Oxford Instruments Ultim Max 40 SDD detector with AZtec software) were conducted to characterize the FGT samples. Elemental compositions were determined via EDX by taking the average of multiple point spectra at several points across the crystals. From these compositions, chemical formulas were calculated by taking ratios of atomic percentages relative to the atomic percentage of Ge, as shown in the Supplementary Materials. X-ray photoelectron spectroscopy (XPS) depth profiles were acquired using a Thermo Fisher Scientific Nexsa G2 Surface Analysis System equipped with a monochromated Al $K_\alpha$ X-ray source (1486.68 eV) and an EX06 monatomic ion source for $Ar^+$ sputtering. X-ray spot sizes of 25 and 50 μm and an $Ar^+$ ion energy of 1000 eV were used. XPS data analysis and peak fitting were performed using Thermo Fisher Scientific's Avantage software, with binding energies charge-corrected to the C 1s aliphatic carbon peak (284.8 eV) following the model developed by Grey et al. [28].

## 3. Results and Discussion

### 3.1. Crystal Structure and Growth Results

Figure 1(a)-(b) show a crystal structure of $Fe_5GeTe_2$ viewed along the *a*- and *c*-axes. The Fe1 and Ge atoms occupy split sites [23], with only one of these configuration shown in Figure 1(a)-(b). The structure in Figure 1(a)-(b) corresponds to the $R3m$ space group (No. 160), where Fe1 atoms occupy sites either above or below the Ge atoms [16]. A more common $Fe_5GeTe_2$ structure belongs to the $R\overline{3}m$ space group (No. 166), in which Fe1 atoms alternate between these sites [16]. $Fe_4GeTe_2$ adopts the same $R\overline{3}m$ structure but lacks Fe atoms at the Fe1 site [13]. In contrast, $Fe_3GeTe_2$ adopts a structure that belongs to the $P6_3/mmc$ space group (No. 194) and possesses only two crystallographically inequivalent

Fe sites: an in-plane Fe site located within the Ge layer (one Fe atom per structural layer) and a second site consisting of two symmetry-equivalent Fe atoms positioned above and below this plane [8]. These structural models are provided as reference for interpreting the compositional data presented below.

Figure 1(c)-(e) shows crystals from Samples 1, 2, and 3, respectively, while Figure 1(f) shows an AFM image demonstrating nanoscale thickness. The crystals exhibit hexagonal or triangular morphologies consistent with the hexagonal FGT crystal structure and prior reports [9,12], with lateral sizes of up to 40 μm and thicknesses as low as 40 nm.

Figure 2(a) shows a representative EDX spectrum for Ni-FGT (Sample 3), and Figure 2(b) shows an SEM image with EDX maps for an undoped FGT crystal (Sample 1). The elemental composition and chemical formula for each sample are shown in Table 1. The calculated chemical formula for the undoped FGT sample (Sample 1) is $Fe_{2.34}GeTe_{1.20}$. To calculate the chemical formulas for the Ni-FGT samples (Samples 2 and 3) are $Fe_{2.43}Ni_{0.10}GeTe_{1.38}$ and $Fe_{3.71}Ni_{0.14}GeTe_{1.50}$, respectively. Taken together, these compositions indicate a consistent deficiency of Fe and Te across all samples. Assuming that the Ni atoms incorporate into the lattice via the Fe-site substitution mechanism reported for Ni-FGT in previous studies [2,25,24,26], this corresponds to chemical formulas of $(Fe_{0.96}Ni_{0.04})_{2.53}GeTe_{1.38}$ and $(Fe_{0.96}Ni_{0.04})_{3.85}GeTe_{1.50}$ for Samples 2 and 3, respectively. The two samples show markedly different total Fe+Ni contents, but a similar Ni-to-Fe ratio with Ni occupying approximately 4% of Fe sites in both cases, demonstrating that our CVD method is capable of producing Ni-FGT with varying Fe concentrations while maintaining consistent levels of Ni dopant incorporation.

### 3.2. Deep-Level XPS Peak Analysis

High-resolution, deep-level XPS (Figures S1-S3 in the Supplementary Materials) were used to determine oxidation states in the etched samples. Except for $Ge^{2+}$ and $Ge^{4+}$, all peak-fitting models included both members of the spin-orbit-split doublets associated with each oxidation state. In the Ge 3d region, $Ge^{0}$ peaks were fitted with a doublet, while other Ge oxidation states were fitted with single peaks, following common practice [29,30,31]. Table 2 lists the fitted binding energies from Figures S1-S3, along with the same energies shifted by fixed offsets; for each doublet, only the lower-energy component is shown.

Considering all four XPS regions together, the energy separations between non-satellite peaks are consistent within each sample, but absolute peak positions show global offsets between samples. Most notably, the  peaks of Sample 2 are systematically redshifted relative to the other two samples by ~0.4-0.6 eV, whereas the peaks of the other two samples differ by ~0.1 eV. These global offsets are attributed to etching-induced effects, as monatomic ion etching at energies of several hundred to thousands of eV can modify surface chemistry through processes such as preferential resputtering, ion implantation, amorphization, and reduction, which can cause uniform binding energy shifts [32].

Using the shifted binding energies in Table 2, the Fe 2p, Ge 3d, and Te 3d spectra across all samples exhibit similar forms consistent with prior reports on FGT [9,29,30,33]. The Fe 2p spectra were fitted following the model used by Nair et al. [9] with $Fe^0$, $Fe^{2+}$, and $Fe^{3+}$ doublets and two associated satellites. $Fe^0$ corresponds to FGT, $Fe^{2+}$ to FeO and/or $FeCO_3$, and $Fe^{3+}$ to $Fe_2O_3$ [33], with $Fe2p_{3/2}$ binding energies of 707.0, 709.8-709.9, and 711.9-712.1 eV, respectively. The Ge 3d region contains $Ge^0$ (FGT), $Ge^{2+}$ (GeO), and $Ge^{4+}$ ($GeO_2$) states [9,14,33] at 29.1-29.2, 31.6, and 32.3-32.4 eV, while the Te 3d region shows $Te^0$ (FGT) and $Te^{4+}$ ($TeO_2$) states [14,29,30,33] with $Te3d_{5/2}$ binding energies at 572.8-572.9 and 576.1-576.3 eV.

In the Ni 2p region, no features are observed for undoped FGT (Sample 1), whereas the Ni-FGT samples (Samples 2 and 3) show two doublets. The lower-energy doublet ($Ni2p_{3/2}$ at 853.0 eV) corresponds to metallic $Ni^0$, while the higher-energy doublet (856.0-856.1 eV) corresponds to $Ni^{2+}$. Although Ni 2p XPS has not previously been reported for Ni-doped FGT, these values are consistent with the reference binding energies for Ni metal (852.6 eV) and $Ni^{2+}$ compounds (853.7-855.6 eV) [34]. The persistence of the Ni signal below the surface confirms Ni is present in the bulk of the material, rather than being confined merely to the surface. Additionally, the presence of $Ni^0$ within the material is consistent with Fe-site substitution as reported in previous studies [2,25,24,26]; however, sputtering-induced chemical reduction is known to be a potential artifact of $Ar^+$ ion sputtering and cannot be ruled out as a contributing factor to the observed $Ni^0$ signal.

### 3.3. XPS Depth Profile

With oxidation states established, we now discuss the XPS depth profile. Figure 3 shows Ni 2p, Fe 2p, Ge 3d, and Te 3d spectra at five depths. Level 0 denotes the native oxidized surface prior to any $Ar^+$ etching, while Levels 1-4 represent sequentially etched depths probing progressively deeper into the material. Notable features are labeled as Peaks 1-8.

Peaks 1 and 2 in Figure 3 are the $Te3d_{5/2}$ components corresponding to $Te^{4+}$ and $Te^0$, respectively. From Level 0 to 1, the $Te^{4+}$ peak decreases while the $Te^0$ peak increases, consistent with the progressive removal of the oxide layer. From Level 2 onward, the $Te^{4+}$ peak is largely absent across all samples.

Peaks 3-6 show similar behavior. Peak 3 comprises Ge3d contributions from $Ge^{2+}$ and $Ge^{4+}$, while Peak 4 corresponds to $Ge^0$. Peak 5 consists of overlapping $Fe2p_{3/2}$ peaks from $Fe^{2+}$ and $Fe^{3+}$, and Peak 6 corresponds to $Fe^0$. In both the Ge 3d and Fe 2p regions, the metallic $Ge^0$ and $Fe^0$ peaks become increasingly dominant with depth. However, the persistence of $Ge^{2+}$, $Ge^{4+}$, $Fe^{2+}$, and $Fe^{3+}$ at Level 4 indicates that the oxide layer is not completely removed.

Peaks 7 and 8 correspond to the $Ni2p_{3/2}$ components of $Ni^{2+}$ and $Ni^0$, respectively. As expected, the undoped FGT sample (Sample 1, Figure 3(a)) shows no Ni signal, while the Ni-FGT samples (Samples 2-3, Figure 3(b)-(c)) exhibit Ni peaks at all depths, demonstrating the presence of Ni within the material. The $Ni^{2+}$ signal is strongest at Level 0 and decreases with depth, whereas the metallic $Ni^0$ signal becomes more prominent at deeper levels. As previously discussed, this depth-dependent behavior is consistent with Ni incorporation at Fe sites within the lattice [2,25,24,26], but does not demonstratively prove this.

## 4. Conclusion

This study demonstrates the viability of chemical vapor deposition for synthesizing Ni-doped FGT nanosheets directly on $SiO_2$/Si substrates with multiple Fe concentrations. In addition to the growth of FGT nanosheets, XPS depth profiling provides clear evidence of Ni in the bulk of the material. In contrast to many other growth methods, this simple and cost-effective approach offers a pathway toward scalable synthesis and supports future research and broader integration of vdW magnets into next-generation spintronic devices.

## 5. Acknowledgements

This work was supported by the startup fund at UC Merced and the Laboratory Directed Research and Development Program of Lawrence Berkeley National Laboratory under U.S. Department of Energy Contract No. DE-AC02-05CH11231. The authors acknowledge funding from NSF awards DGE-2125510 and DMR-2425230. The authors are sincerely grateful for the service provided by the IMF facility at UC Merced.

## 6. Declaration of Generative AI Use

During the preparation of this work, the authors used Gemini 3 Pro to create the experimental setup diagram (Figure 1(g)) and ChatGPT 5 to initially draft the abstract and conclusion and to condense and refine the other sections of this manuscript. After using these tools, the authors reviewed and edited the content as needed and take full responsibility for the content of the published article.

## 7. Data Availability

Data will be made available upon reasonable request.

## 8. Credit Authorship Contribution Statement

**Matthew Metcalf**: Data curation, formal analysis, investigation, methodology, resources, supervision, validation, visualization, writing – original draft, writing – review & editing.

**Jesse Martinez**: Investigation.

**Armella Mushfique**: Investigation.

**Alexander Riou**: Investigation.

**Lutfun Nahar**: Investigation, writing – review & editing.

**Bamidele Onipede**: Investigation, resources, writing – review & editing.

**Hui Cai**: Conceptualization, funding acquisition, methodology, resources, supervision, writing – review & editing.

Table 1. Elemental Compositions.

| | Ni At% | Fe At% | Ge At% | Te At% | Chemical Formula |
|---|---|---|---|---|---|
| Sample 1 | 0.0% | 51.5% | 22.0% | 26.5% | $Fe_{2.34}GeTe_{1.20}$ |
| Sample 2 | 2.0% | 49.5% | 20.4% | 28.1% | $(Fe_{0.96}Ni_{0.04})_{2.53}GeTe_{1.38}$ |
| Sample 3 | 2.2% | 58.4% | 15.7% | 23.6% | $(Fe_{0.96}Ni_{0.04})_{3.85}GeTe_{1.50}$ |

Average elemental compositions of samples.

Table 2. XPS Peak Positions.

| Peak | Sample 1 | Sample 2 | Sample 3 | Sample 1 + 0.1 eV | Sample 2 + 0.5 eV | Sample 3 + 0.0 eV |
|---|---|---|---|---|---|---|
| $Ni2p_{3/2}$ $Ni^0$ | — | 852.5 | 853.0 | — | 853.0 | 853.0 |
| $Ni2p_{3/2}$ $Ni^{2+}$ | — | 855.6 | 856.0 | — | 856.1 | 856.0 |
| $Fe2p_{3/2}$ $Fe^0$ | 706.9 | 706.5 | 707.0 | 707.0 | 707.0 | 707.0 |
| $Fe2p_{3/2}$ $Fe^{2+}$ | 709.7 | 709.4 | 709.9 | 709.8 | 709.9 | 709.9 |
| $Fe2p_{3/2}$ $Fe^{3+}$ | 712.0 | 711.4 | 711.9 | 712.1 | 711.9 | 711.9 |
| $Fe2p_{3/2}$ $Fe^{2+}$ Satellite | 715.7 | 715.1 | 715.3 | 715.8 | 715.6 | 715.3 |
| $Fe2p_{3/2}$ $Fe^{3+}$ Satellite | 720.0 | 719.4 | 720.0 | 720.1 | 719.9 | 720.0 |
| $Ge3d_{5/2}$ $Ge^0$ | 29.0 | 28.7 | 29.1 | 29.1 | 29.2 | 29.1 |
| Ge3d $Ge^{2+}$ | 31.5 | 31.1 | 31.6 | 31.6 | 31.6 | 31.6 |
| Ge3d $Ge^{4+}$ | 32.3 | 31.9 | 32.3 | 32.4 | 32.4 | 32.3 |
| $Te3d_{5/2}$ $Te^0$ | 572.8 | 572.3 | 572.9 | 572.9 | 572.8 | 572.9 |
| $Te3d_{5/2}$ $Te^{4+}$ | 576.2 | 575.6 | 576.2 | 576.3 | 576.1 | 576.2 |

Peak identities (first column) with raw binding energies (next three columns) and shifted binding energies (last three columns) after correcting for global offsets.

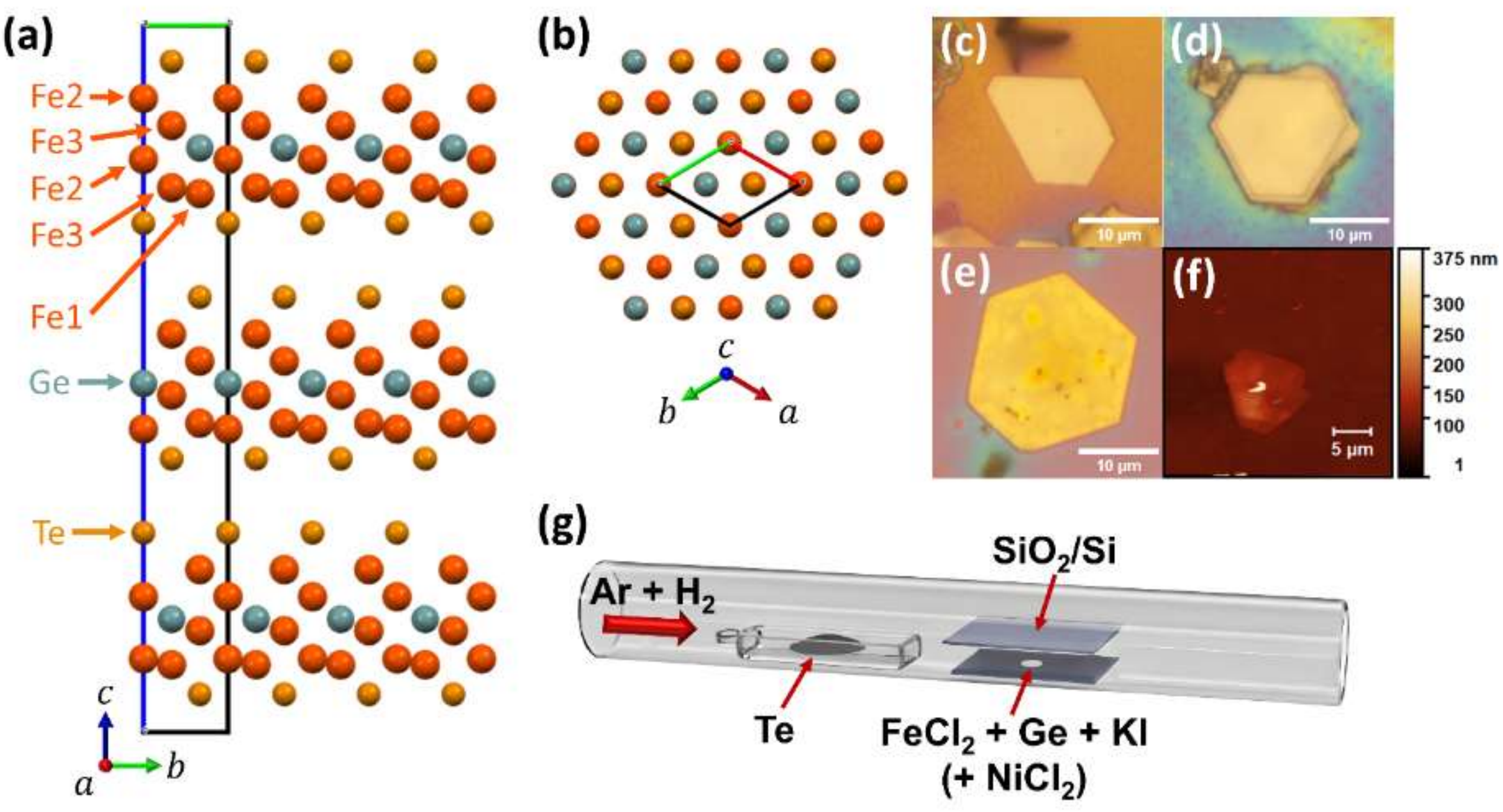

Figure 1. Crystal structure, crystal images, and experimental setup. (a)-(b) $Fe_5GeTe_2$ crystal structure and unit cells ($R3m$, no. 160). Generated from ICSD Collection Code 130074 [24]. (c)-(e) Crystals from the growths of Samples 1-3, respectively. (f) AFM image showing nanoscale thickness of 40-100 nm. (g) Diagram of CVD setup.

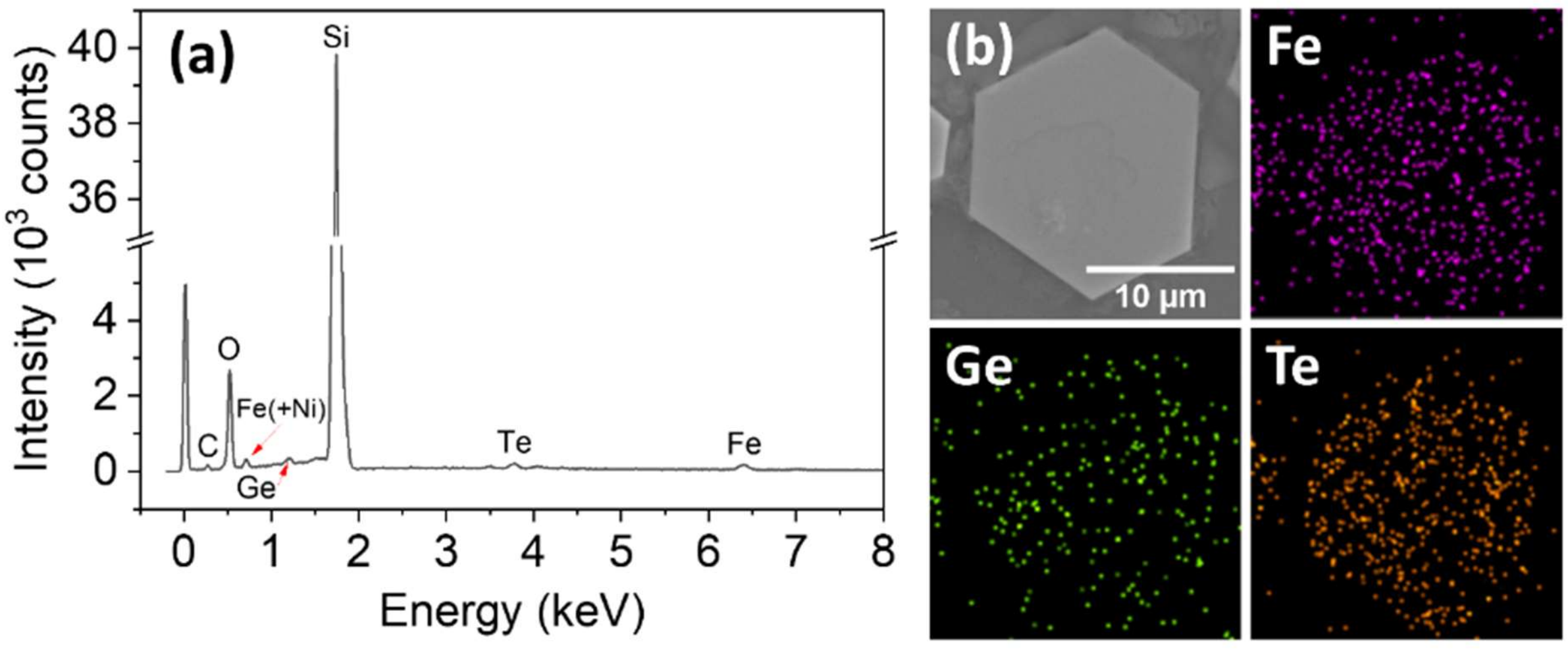

Figure 2. EDX characterization. (a) Representative EDX spectrum of Ni-FGT (Sample 3). (b) SEM image and corresponding EDX elemental maps of undoped FGT.

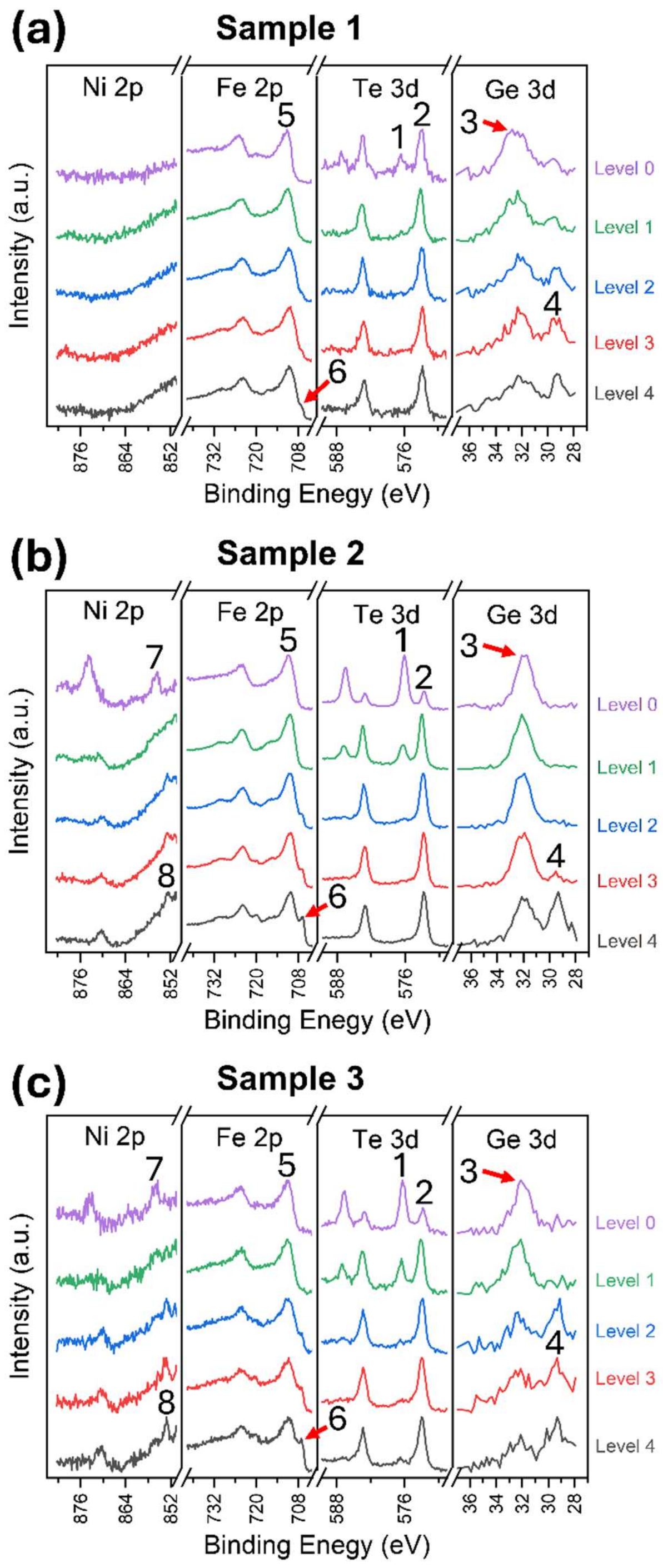

Figure 3. XPS depth profiles with labeled peaks. (a) Sample 1 (Undoped FGT). (b) Sample 2 (Ni-FGT). (c) Sample 3 (Ni-FGT with more Fe). Peaks 1 and 2 demonstrate a shift from predominantly $Te^{4+}$ in $TeO_2$ to $Te^0$ in FGT. Peaks 3 and 4 show a similar shift from $Ge^{4+}$ in $GeO_2$ to $Ge^0$ in FGT. Peak 5 is combination of $Fe^{2+}$ and $Fe^{3+}$ peaks that correspond to a combination of FeO, $FeCO_3$, and $Fe_2O_3$. At deep etch levels, Peak 6 emerges and corresponds to $Fe^0$ in FGT. Peaks 7 and 8 demonstrate a shift from predominantly $Ni^{2+}$ (likely NiO) to $Ni^0$.

## Supplementary Material

# Chemical Vapor Deposition of Ni-doped Iron Germanium Telluride Nanosheets

Matthew Metcalf[a], Jesse Martinez[a], Armella Mushfique[a], Alexander Riou[a], Lutfun Nahar[b], Bamidele Onipede[a], Hui Cai[a,c,*]

[a] Department of Physics, University of California, Merced, Merced CA 95343, United States of America

[b] Department of Physics, Arizona State University, Tempe AZ 85281, United States of America

[c] Materials Sciences Division, Lawrence Berkeley National Laboratory, Berkeley CA 94720, United States of America

* Corresponding author: Hui Cai (hcai6@ucmerced.edu)

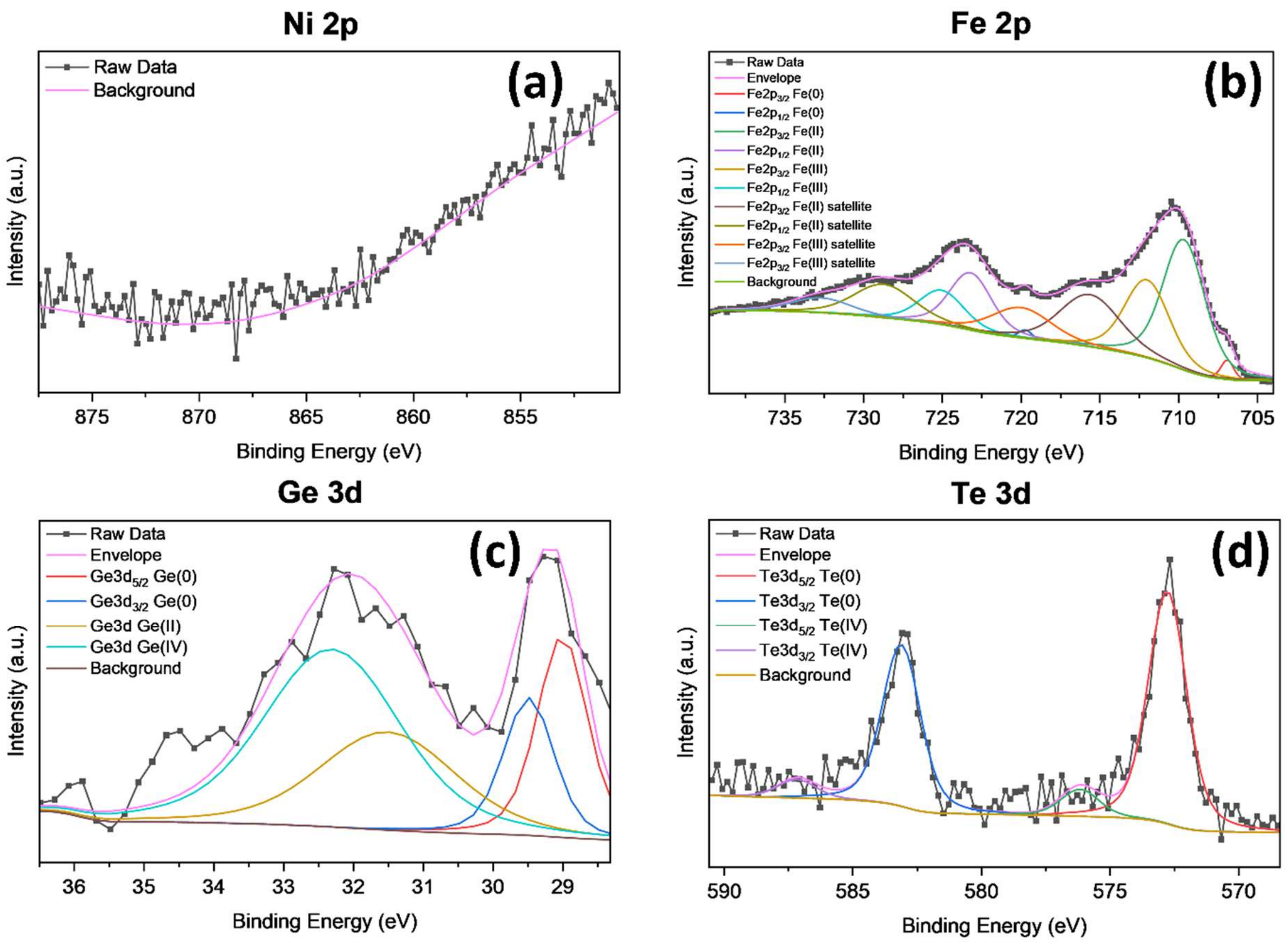

Figure S1. Deep-level XPS of Sample 1 (undoped FGT) after etching. (a) Ni 2p region shows no features above the background. (b) Fe 2p region fitted with five-doublet model used by Nair et al. [1] and shows $Fe^0$, $Fe^{2+}$, and $Fe^{3+}$ oxidation states. (c) Ge 3d region shows $Ge^0$, $Ge^{2+}$, and $Ge^{4+}$ oxidation states. (d) Te 3d region shows $Te^0$ and $Te^{4+}$ oxidation states.

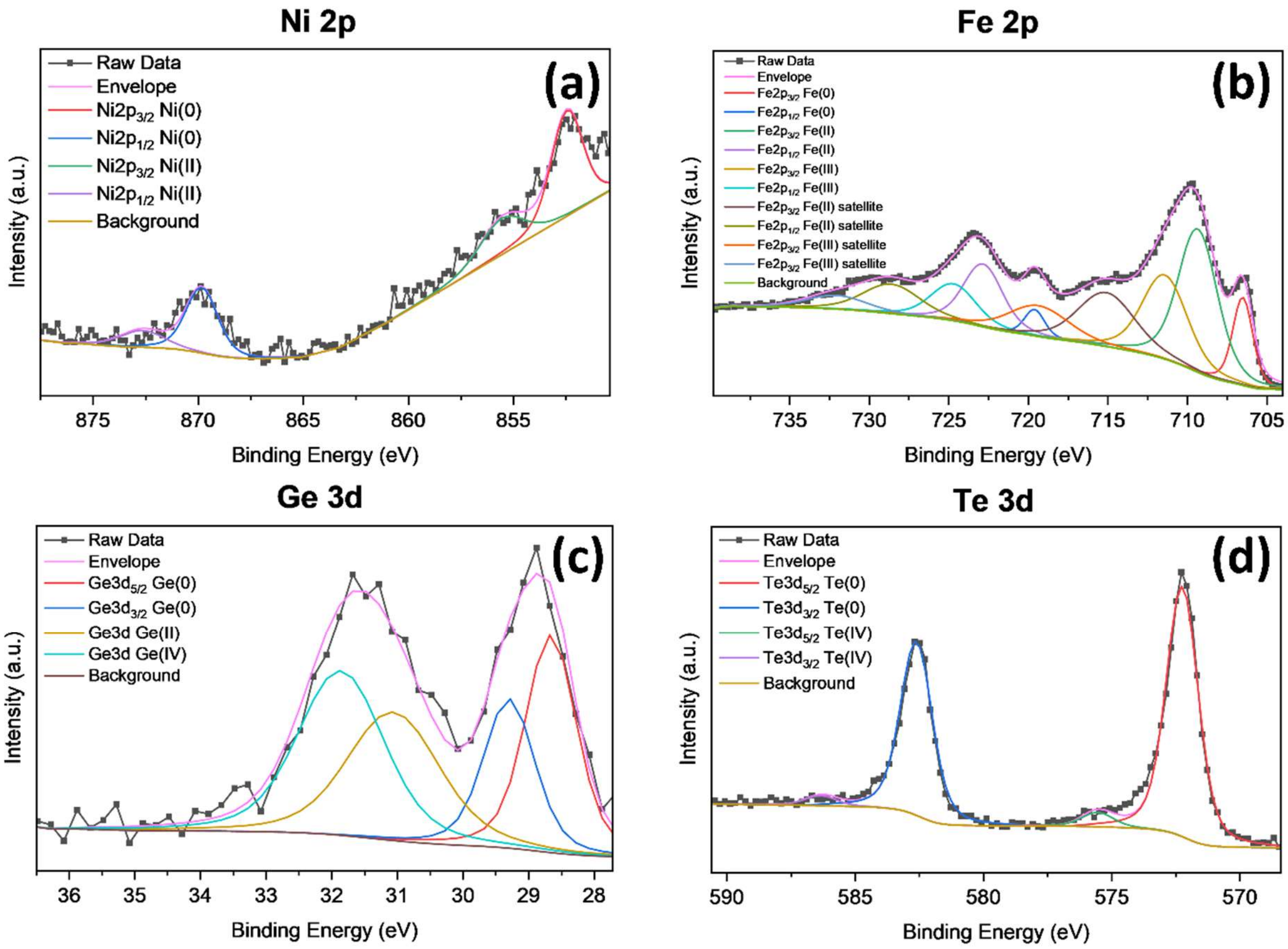

Figure S2. Deep-level XPS of Sample 2 (Ni-FGT) after etching. (a) Ni 2p region shows two doublets corresponding to $Ni^0$ and $Ni^{2+}$ oxidation states. (b) Fe 2p region fitted with five-doublet model used by Nair et al. [1] and shows $Fe^0$, $Fe^{2+}$, and $Fe^{3+}$ oxidation states. (c) Ge 3d region shows $Ge^0$, $Ge^{2+}$, and $Ge^{4+}$ oxidation states. (d) Te 3d region shows $Te^0$ and $Te^{4+}$ oxidation states.

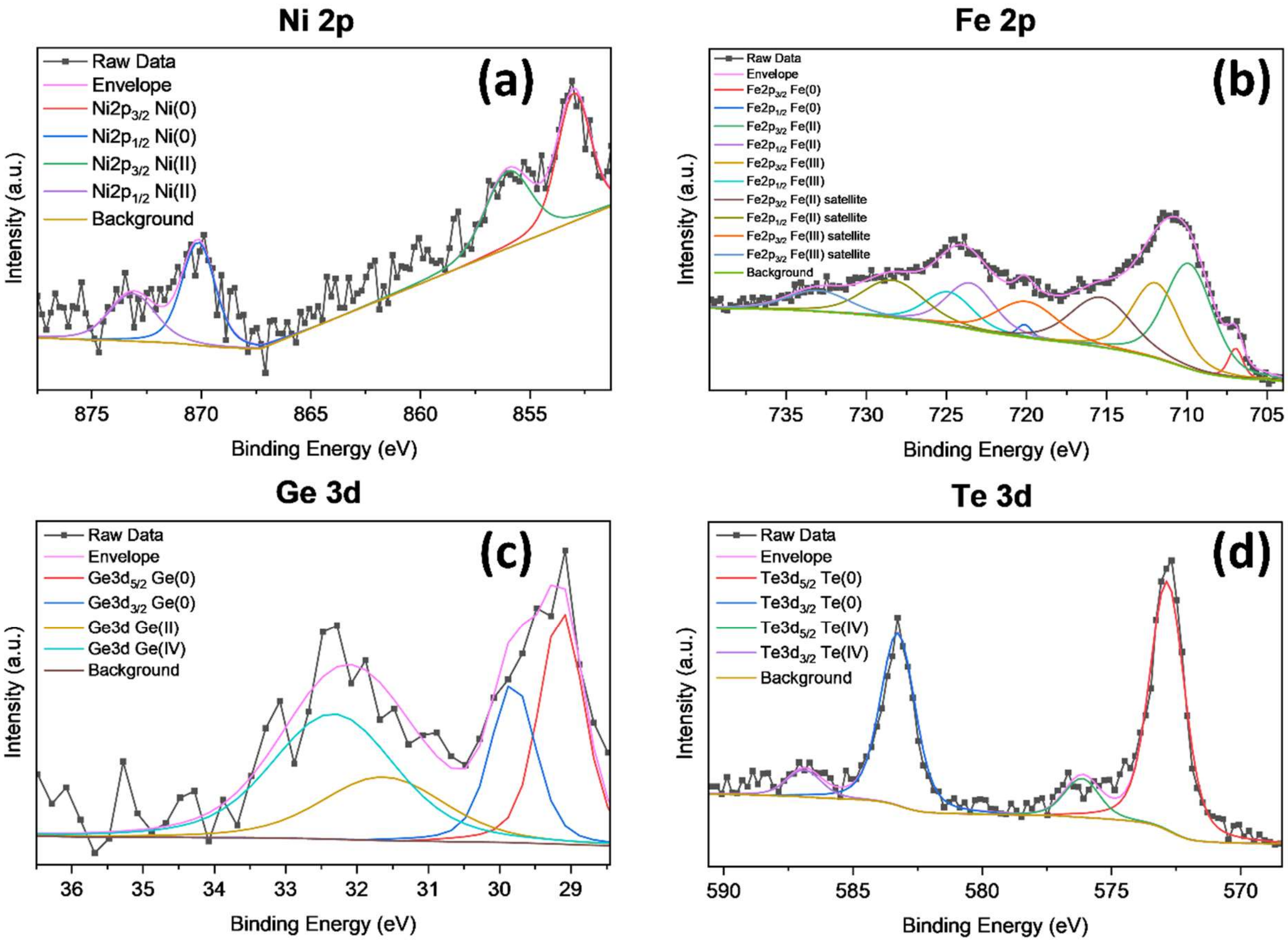

Figure S3. Deep-level XPS of Sample 3 (Ni-FGT with higher Fe content) after etching. (a) Ni 2p region shows two doublets corresponding to $Ni^{0}$ and $Ni^{2+}$ oxidation states. (b) Fe 2p region fitted with five-doublet model used by Nair et al. [1] and shows $Fe^{0}$, $Fe^{2+}$, and $Fe^{3+}$ oxidation states. (c) Ge 3d region shows $Ge^{0}$, $Ge^{2+}$, and $Ge^{4+}$ oxidation states. (d) Te 3d region shows $Te^{0}$ and $Te^{4+}$ oxidation states.